\documentclass[proof]{WileyASNA-v1}

\articletype{Article Type}%

\received{}
\revised{}
\accepted{}

\begin{document}

\title{Quasars as high-redshift standard candles}

\author[1,2]{G. Risaliti*}
\author[1,2]{E. Lusso}
\author[2]{E. Nardini}
\author[3,4]{G. Bargiacchi}
\author[5]{S. Bisogni}
\author[6]{A. Sacchi}
\author[1,2]{M. Signorini}
\author[1,2]{B. Trefoloni}

\authormark{G. Risaliti}

\address[1]{Department of Physics and Astronomy, University of Florence, Via G. Sansone 1, 50019, Sesto Fiorentino (FI), Italy}

\address[2]{INAF-Arcetri Astrophysical Observatory, Largo E. Fermi 5, 50125 Firenze, Italy}

\address[3]{Scuola Superiore Meridionale, Largo S. Marcellino 10, 80138 Napoli, Italy}

\address[4]{INFN – Sezione di Napoli, Via Cinthia 9, 80126 Napoli, Italy}

\address[5]{INAF – Istituto di Astrofisica Spaziale e Fisica Cosmica Milano, Via Corti 12, 20133 Milano, Italy}

\address[6]{Istituto Universitario di Studi Superiori (IUSS), Piazza della Vittoria 15, 27100 Pavia, Italy}

\corres{*\email{guido.risaliti@unifi.it}}


\abstract{In the past few years, we built a Hubble diagram of quasars up to redshift z$\sim$7, based on the non-linear relation between quasars' X-ray and UV luminosities. Such a Hubble diagram shows a $>4 \sigma$ deviation from the standard flat $\Lambda$CDM model at $z>1.5$. Given the important consequences of this result, it is fundamental to rule out any systematic effect in the selection of the sample and/or in the flux measurements, and to investigate possible redshift dependences of the relation, that would invalidate the use of quasars as standard candles. Here we review all the observational results supporting our method: the match of the Hubble diagram of quasars with that of supernovae in the common redshift range, the constant slope of the relation at all redshifts, the redshift non-evolution of the spectral properties of our sources both in the X-rays and in the UV. An independent  test of our results requires the observation of other standard candles at high redshift. In particular, we expect that future observations of supernovae at $z>2$ will confirm the deviation from the concordance model found with the Hubble diagram of quasars.}



\maketitle


\section{Introduction and Selection}

The non-linear relation between X-ray and UV luminosities in quasars is usually described as:
\begin{equation}
log(L_X)= \beta+\alpha\times log(L_{UV})
\end{equation}

where $L_X$ and $L_{UV}$ are the monochromatic luminosities at 2~keV and 2500~\AA, respectively, and $\alpha\sim0.6$ (e.g. \citealt{Lusso+20}, \citealt{Bisogni+21}). This empirical relation must be the manifestation of the physical mechanism responsible for the energy flow from the accretion disk to the hot X-ray corona. Even if we know that this link between disk and corona must exist and be stable (otherwise the corona would cool down very rapidly) no complete physical model reproducing the relation has been proposed so far. As a consequence, the validation of the relation as a cosmological probe can be obtained only through an observational approach. 

The sample used to build the Hubble diagram of quasars consists of 2,420 sources (\citealt{Lusso+20}) drawn from a parent sample of more than 13,000 quasars with UV and X-ray spectral information, mostly obtained by cross-correlating the SDSS quasar catalogue DR14 (\citealt{Paris+18}) with the 4XMM-DR9 catalogue of serendipitous XMM-Newton observations (\citealt{Webb+20}) and with the Chandra Second Point Source Catalogue (\citealt{Evans2010}). This means that our filters removed about 80\% of the sources. Therefore, the first obvious problem to consider is the possible presence of a redshift-dependent systematic effect in the selection procedure. 
We applied three main filters:\\
1) "Eddington bias": in order to avoid brighter-than-average states, we accepted only objects with a deep enough X-ray observation to be detected even if they are caught in a relatively low X-ray state. This procedure is based on an {\it expected} X-ray flux, so in principle it could  introduce a bias; however we have demonstrated (\citealt{Lusso+20}) that the effect of this possible systematic is negligible, provided that we choose a conservative threshold for the depth of the observation, at the price of reducing the sample statistics.\\
2) X-ray spectrum: we selected quasars with an X-ray photon index $\Gamma>1.7$, in order to avoid gas-obscured objects, for which we may derive an incorrect X-ray flux.\\
3) Optical spectrum: we selected relatively blue objects (using the standard quasar SED of \cite{VandenBerk01} as a reference). This helps to remove objects with a significant dust extinction.\\

These filters are based on the quality of the observations and on the accuracy of the flux measurements, and none of them are directly related to the physical properties of quasars. However, the second and third one, besides excluding objects with either gas absorption and/or dust extinction, may also filter out quasars with a non-standard emission, such as an intrinsically flat  X-ray spectrum, or and intrinsically red optically spectrum. Even if this is the case for some objects, it does not affect the cosmological application: it only means that our study is not conducted on the whole quasar population, but only on "standard" X-ray steep, optically blue objects. Finally, we notice that the precise choice of the threshold values for the X-ray slope and the optical colour does not affect the final outcome (\citealt{Lusso+20}).

\section{The X-ray to UV relation in quasars}

Since we want to use the non linearity of the relation to derive quasar distances, it does not make any sense to analyse it in luminosity units, if not just for illustration purposes, as in \citealt{RL19}. Instead, it is possible to split the sample in small redshift intervals, in order to neglect the distance differences among sources in the same interval. We found that, given the observed dispersion, a bin size of $\Delta\log(z)$$\sim$0.1 fulfills this requirement, and allows a cosmology-independent analysis of the relation. With this bin size, we divided the sample in 12 intervals, from $z$=0.4 to $z$=5.0. We show in Fig.~1 the results of the analysis for three random redshift intervals. The complete analysis in shown in (\citealt{Lusso+20}). Both the visual inspection of the figure and the numerical results clearly reveal that the relation holds at all redshifts and does not show any significant deviation from linearity. Furthermore, the slope is consistent with being constant at all redshifts. In this scenario, the cosmological information is encoded in the normalization of the relation: if the {\it physical} relation between X-ray and UV luminosities is that in Equation~1, and the parameter $\beta$ is constant with redshift, then the flux-flux relation in a small redshift bin is:

\begin{equation}
	log(F_X)\sim \beta^*(z)+\alpha\times log(F_{UV})
\end{equation}

where the slope $\alpha$ is constant, and is the same as in Equation~1, while the parameter $\beta^*$ depends on the source distance (hence on the cosmological model):
\begin{equation}
	\beta^*=\beta+(\alpha-1)\log(4\pi)+2(\alpha-1)\log(D_L)
\end{equation}

where $D_L$ is the luminosity distance.

\begin{figure}[t]
\centerline{\includegraphics[width=9cm]{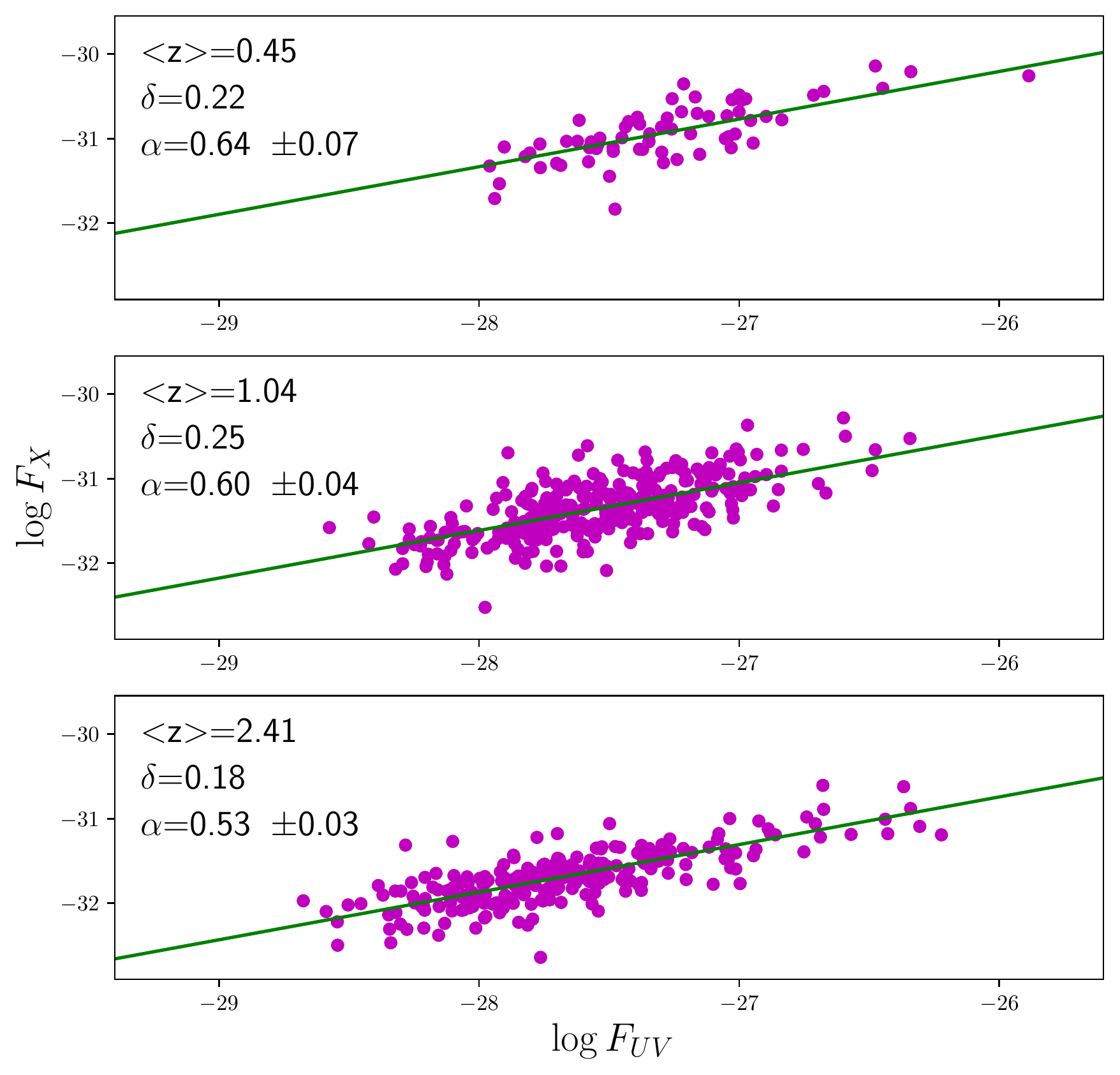}}
	\caption{Examples of the analysis of the X-ray to UV relation in flux units, in three small redshift bins. $\alpha$ is the slope of the best fit relation (Equation~2), while $\delta$ is the intrinsic dispersion of the relation.\label{fig1}}
\end{figure}

In Fig.~2 we show the $\beta^*$-redshift relation, fitted with a cosmographic model, as described in \citet{Bargiacchi+21}, and compared with a $\Lambda$CDM model with $\Omega_M$=0.3, $\Omega_\Lambda$=0.7. This diagram contains the same physical information as a Hubble diagram. In fact, {\it it is} a Hubble diagram where each point represents the average distance of the objects in the relative redshift interval. The tension with the $\Lambda$CDM model is apparent from this plot, and has been more quantitatively measured in \citet{Bargiacchi+21}. 

The significance of this tension relies on a hypothesis (the non-evolution of the relation) that  cannot be directly probed at high redshifts (we would need another standard candle at those redshifts). However, it is possible to test it at low redshifts ($z$$<$1.5) where supernovae are available. In Fig.~3 we show a Hubble diagram of supernovae and quasars in the $z$=0--1.5 range. In this plot, the only degree of freedom for quasars is a global cross-calibration parameter, while the shape of the diagram is only determined by the X-ray to UV flux relation. The perfect match between the two indicators is a direct proof of a non-evolution of the relation up to $z$$\sim$1.5.

\begin{figure}[t]
\centerline{\includegraphics[width=9cm]{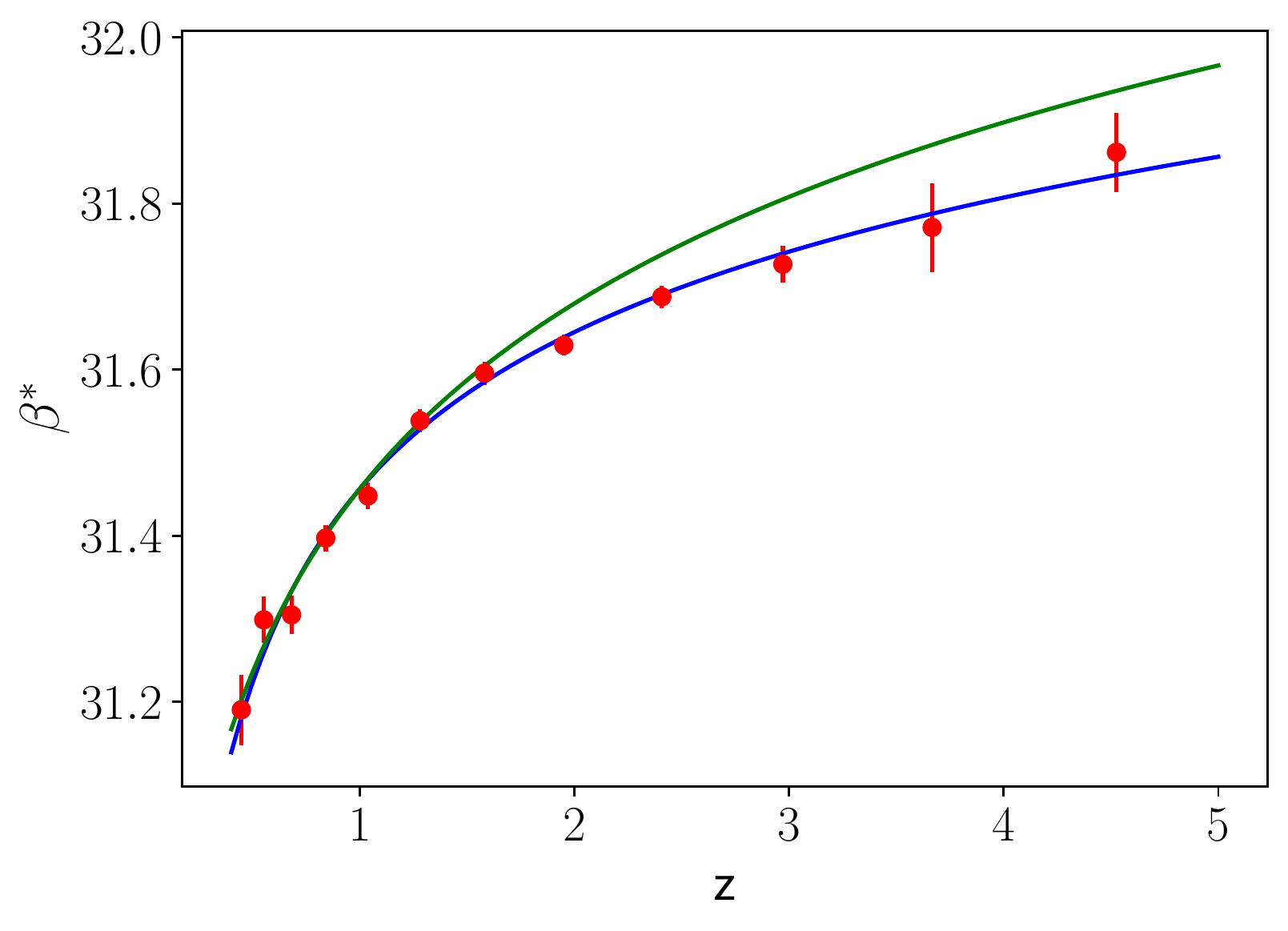}}
	\caption{Relation between the parameter $\beta^*$ in Eq.~2 and the redshift. Since the information on the distance is encoded in $\beta^*$, this plot is equivalent to a standard Hubble diagram. The two curves represent a best-fit cosmographic model (blue line) and a flat $\Lambda$CDM model with $\Omega_M$=0.3.\label{fig2}}
\end{figure}

\begin{figure}[t]
\centerline{\includegraphics[width=9cm]{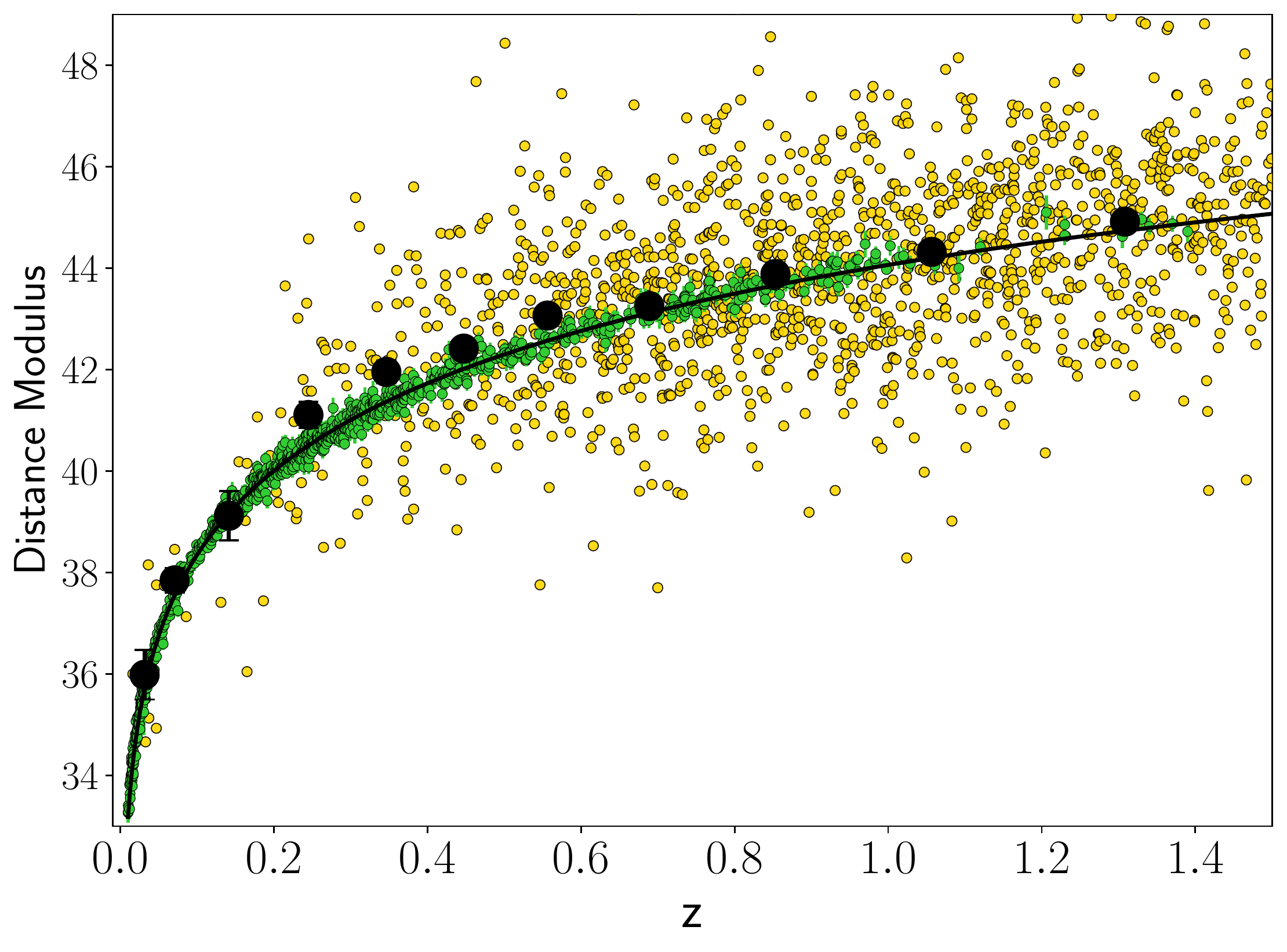}}
	\caption{Hubble diagram of supernovae (green points) and quasars (yellow points) in the common redshift range. Black dots are quasar averages in small redshift bins.\label{fig2}}
\end{figure}

\section{Origin of the dispersion}

In the previous Section we have shown that the X-ray to UV luminosity relation has a constant slope at all redshift, and a constant normalization up to z$\sim$1.5. It is therefore reasonable to assume that it remains constant at higher redshifts too. However, the dispersion of the relation (and, as a consequence, of the Hubble diagram) is quite high: the total dispersion is of the order of 0.25~dex, which means that the uncertainty of the distance estimate of an individual quasar is as high as 80\%. Most of this dispersion is not due to the statistical errors in the X.ray and UV flux measurements, that on average contribute by only $\sim$0.05~dex: once the measurements errors are taken into account, we are left with a $\sim$0.22~dex {\em} intrinsic dispersion. This high value is a strong limitation of our method for two reasons: it reduces the statistical significance of the cosmological constraints, and it makes it more difficult to rule out possible hidden systematic effects. For this reason it is fundamental to (a) understand the origin of this dispersion, and (b) if possible, reduce it.

The first point will be treated in detail in a forthcoming publication (Signorini et al., in prep.). Here we only notice that there are at least two "external" factors significantly contributing to the dispersion: variability (UV and X-ray observations are in most cases not simultaneous, and even if they were, the different light travel times would randomize the measured X-ray to UV ratio) and disk inclination (assuming an isotropic X-ray emission). 

Regarding the remaining dispersion, in a paper recently published by our group (\citealt{Sacchi+22}) we show that if we select a "golden" subsample with high-quality X-ray observations and we perform a one-by.one spectroscopic analysis, the dispersion drops to values of the order of 0.1~dex. In particular, we considered a sample of 30 objects in the z=3.0-3.3 range with very high spectral quality (many X-ray observations are pointed, rather than serendipitous), and we obtained the results shown in Fig.~4: the slope is still  the same as at low redshift, and the intrinsic  dispersion is only 0.09~dex. Such a small value is entirely accounted for by variability and inclination effects. As a consequence, the real intrinsic dispersion of the relation must be  close to zero.

\begin{figure}[t]
\centerline{\includegraphics[width=9cm]{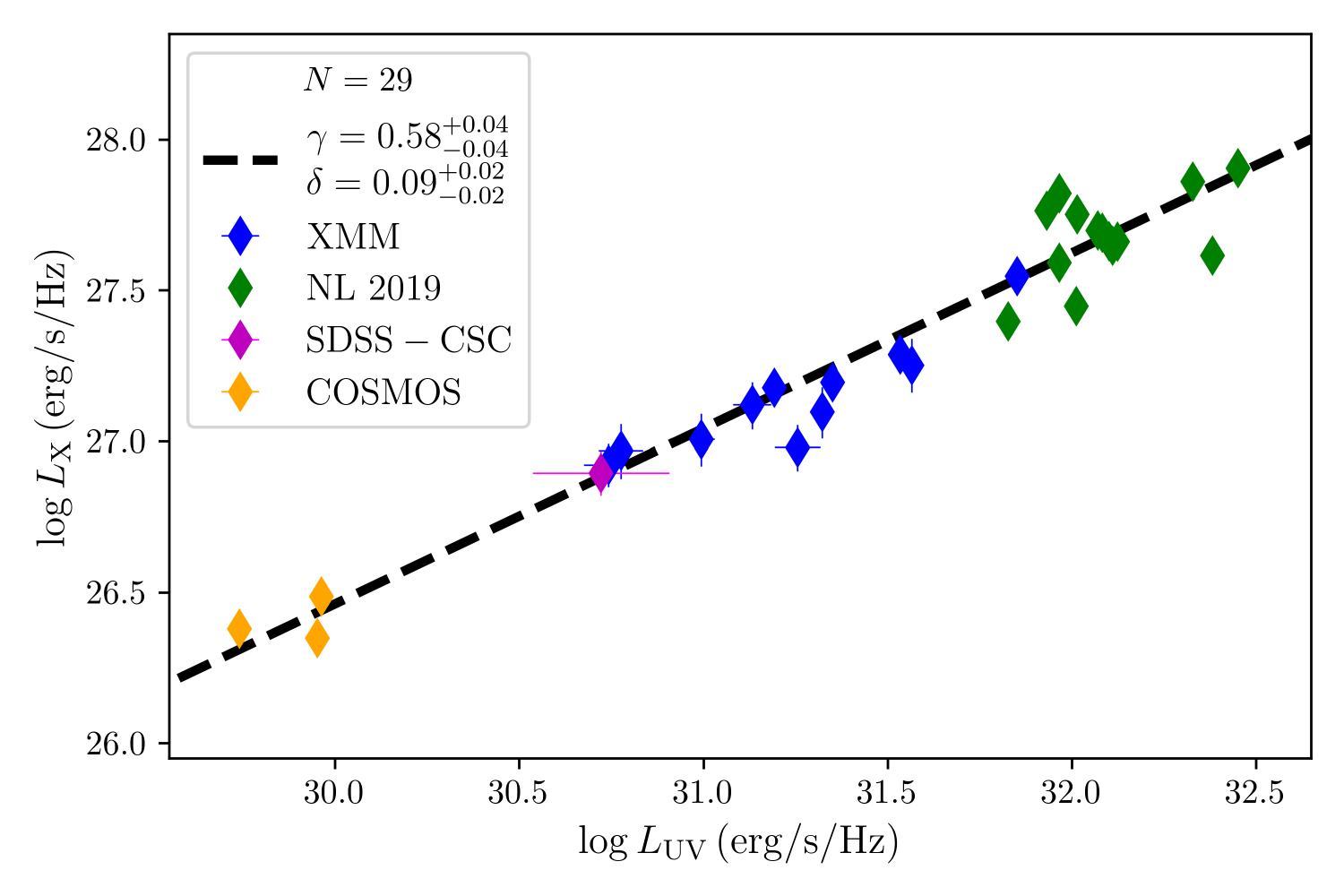}}
	\caption{X-ray to UV relation for the "golden" sample at z=3.0-3.3. Green points (labeled "NL 2019") represent pointed XMM-Newton observations (\citealt{Nardini+19}); yellow points (labeled "COSMOS") are Chandra-Cosmos sources (Bisogni et al.~2021); blue points (labeled "XMM") are XMM-Newton serendipitous sources with a complete spectroscopic analysis (Sacchi et al.~2022); the purple point (labeled "SDSS - CSC") is a serendipitous Chandra source (Bisogni et al.~2021).\label{fig4}}
\end{figure}

\section{Spectral properties of high-redshift quasars}\label{sec2}

A final, fundamental check on the reliability of the relation concerns the spectroscopic properties of the sources, in order to rule out the possibility that the low observed dispersion in the "golden" sample is due to the selection of peculiar objects. 
Fig.~5 shows the stacked X-ray and optical (rest-frame UV) spectra of the "golden" sample, compared with standard references: a power law with slope $\Gamma=1.9$ in the X-rays, and the (\citealt{VandenBerk01}) quasar composite spectrum. It is clear that the sources in our sample are by no means different from standard quasars at lower redshift. This is a further indication of a likely constancy of the relation with redshift. 

\begin{figure}[t]
\centerline{\includegraphics[width=9cm]{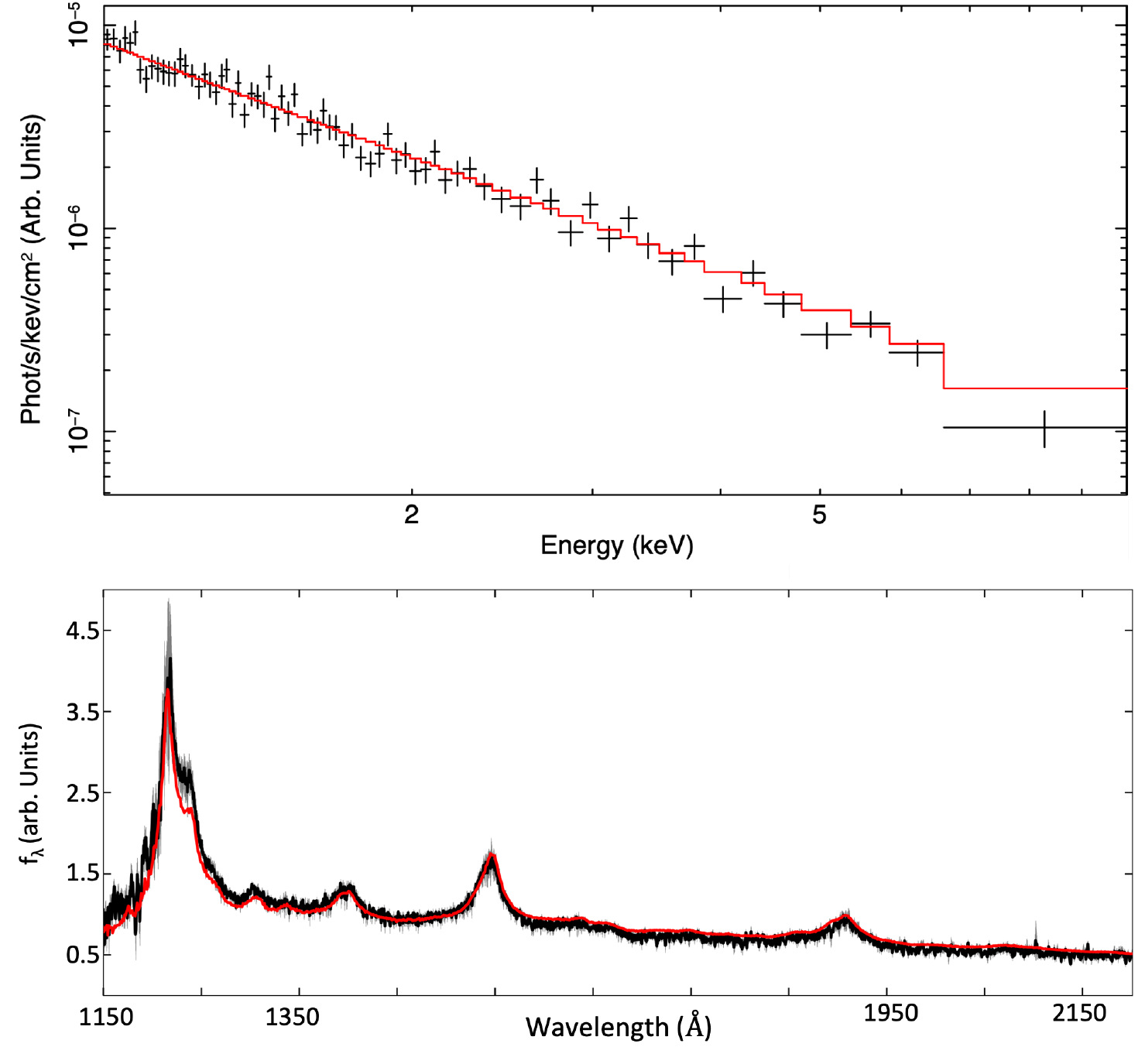}}
	\caption{From \citet{Sacchi+22}. Top panel: average X-ray spectrum for our z$\sim$3 sample, compared with a power-law model with photon index $\Gamma=1.9$. Bottom panel: average spectral energy distribution in the rest-frame UV for the same sample, compared with the average SED of \citet{VandenBerk01}, obtained from a sample of several thousands of quasars in a wide range of redshifts and luminosities).\label{fig5}}
\end{figure}

\section{Discussion and Conclusions}

In this paper we discussed the reliability of quasars as high-redshift standard candles through the non linear X-ray to UV luminosity relation.

Several observational results suggest that this relation holds at all redshifts and luminosities with little, if any, intrinsic dispersion:\\
1) The analysis of the relation in small redshift bins shows, in a cosmology-independent way, that its slope is constant with redshift.\\
2) The Hubble diagram of quasars and supernovae are in nearly perfect agreement in the common redshift range, z$\sim$0.-1.5.\\
3) The intrinsic dispersion of the relation is as small as 0.09 dex for a "golden" sample with particularly reliable X-ray flux measurements. Such small dispersion can be entirely accounted for by variability and disk inclination effects.\\
4) There is no  difference in either the X-ray or the UV between the "golden" sample spectra  and the average quasar spectra from large, lower redshift samples.

These results strongly support the two main conclusions of our work:\\
1) The X-ray to UV relation in quasars is the consequence of a universal physical mechanism regulating the energy transfer from the accretion disk to the hot corona, which holds at all redshifts and luminosities.\\
2) We have convincing evidence supporting the adoption of quasars as standard candles beyond the maximum redshift directly testable with independent distance indicators (i.e., supernovae up to z$\sim$1.5).

Given the absence of a general physical model reproducing the relation, the only way to further test our results is through new standard candles at higher redshift. We expect that future detections of supernovae at $z>1.5$ will confirm the deviations from the standard cosmological model found with the Hubble diagram of quasars. 


\section*{Acknowledgments}

GR and EL acknowledges the support of grant ID: 45780 Fondazione Cassa di Risparmio Firenze.

\bibliography{Risaliti-XMM2022}%

\end{document}